\documentstyle{pss}
\input epsf.tex
\begin{document}

\title{Non-Ergodicity of the 1D Heisenberg Model}

\author{{\sc M. Bak}\footnote{Corresponding author;
Phone: +39 089 965418, Fax: +39 089 965275, E-mail:
karen@delta.amu.edu.pl. On leave from Institute of Physics, Adam
Mickiewicz University, Poznan, Poland.}, {\sc A. Avella} and {\sc
F. Mancini}}

\address{Dipartimento di Fisica ``E.R. Caianiello'' - Unit\`a INFM di Salerno\\
Universit\`a degli Studi di Salerno, I-84081 Baronissi (SA),
Italy}

\submitted{May 10, 2002} \maketitle

\hspace{9mm} Subject classification: 05.30.-d; 71.10.-w

\begin{abstract}
The relevance of zero-energy functions, coming from zero-energy
modes and present in the structure of bosonic Green's functions,
is often underestimated. Usually, their values are fixed by
assuming the ergodicity of the dynamics, but it can be shown that
this is not always correct. As the zero-energy functions are
connected to fundamental response properties of the system under
analysis (specific heat, compressibility, susceptibility, etc.),
their correct determination is not an irrelevant issue. In this
paper we present some results regarding the zero-energy functions
for the Heisenberg chain of spin-1/2 with periodic boundary
conditions as functions of the number of sites, temperature and
magnetic field. Calculations are pursued for finite chains, using
equations of motion, exact diagonalization and Lanczos technique,
and the extrapolation to thermodynamic limit is studied.

\end{abstract}

\section{Introduction}

The issue of ergodicity of physical systems is not new. A system
is ergodic if it goes through every point of phase space during
its time evolution. In such systems, the equilibrium averages
(i.e., the time averages) are equal to the ensemble averages,
which are much easier to compute. Anyway, we have to face the
problem of non-ergodicity of many physical systems (e.g., the
existence of even one integral of motion divides the phase space
into separate subspaces not connected by the dynamics). The lack
of ergodicity has measurable effects as the well-known difference
between the static isolated (or Kubo \cite{kubo}) susceptibility
and the isothermal one \cite{suzuki,suzukiB}.

In the widely used formalism of Green's functions the issue of
ergodicity appears as a difficulty in the determination of the
zero-frequency functions present in the bosonic propagators
%\cite{stevens,fernandez,callen,lucas,morita,kwok,ramos,fick}.
\cite{stevens}-\cite{fick}. The problem reappears also in e.g.,
composite operator method \cite{mancini}. The equations of motion
do not uniquely determine causal Green's functions and correlation
functions but only up to some momentum function which severely
affects the self-consistent calculation of the retarded Green's
functions too. Usually, these zero-frequency functions are fixed
by assigning them their ergodic values, but this can not be
justified a priori. Wrong determination of them dramatically
affects the values of directly measurable quantities like
compressibility, specific heat, magnetic susceptibility. According
to this, they should be calculated case by case.

In this paper, we calculate the zero-frequency functions for the
Heisenberg chain of spin-1/2 and show that they take non-ergodic
values for finite lengths and in the bulk limit too.

\section{Definitions}

One of the main aims of condensed matter calculations is to
compute the correlation functions of the physical system under
analysis:
\begin{equation}
C(i,j)=\left\langle\psi(i)\psi^{\dagger}(j)\right\rangle
\end{equation}
where $i=({\bf i},t)$ stands for both spatial ${\bf i}$ and
temporal $t$ coordinates and $\psi$ for any operator relevant to
the dynamics under study. The Fourier transform in frequency of
the correlation functions can be expressed in terms of the
eigenstates $| n\rangle$ and eigenenergies $E_{n}$ of the system
in the following way
\begin{eqnarray}
C({\bf i},{\bf j},\omega) &=& 2\pi\delta(\omega) \Gamma_{\psi\psi^{\dagger}} ({\bf i},{\bf j}) \nonumber \\
&+& {2\pi \over Z} \sum_{ \stackrel{n,m}{E_{n} \neq E_{m}}} {\rm
e}^{-\beta E_{n}} \langle n | \psi({\bf i}) | m \rangle \langle m
| \psi^{\dagger}({\bf j}) | n \rangle \delta \left[ \omega +
\left( E_{n} - E_{m} \right) \right]
\end{eqnarray}
where the {\em zero-frequency function}
$\Gamma_{\psi\psi^{\dagger}}({\bf i},{\bf j})$ is defined as:
\begin{equation}\label{Gdef}
\Gamma_{\psi\psi^{\dagger}}({\bf i},{\bf j})={1\over
Z}\sum_{\stackrel{n,m}{E_{n}=E_{m}}} {\rm e}^{-\beta E_{n}}
\langle n | \psi({\bf i}) | m \rangle \langle m |
\psi^{\dagger}({\bf j}) | n \rangle
\end{equation}
Its ergodic value is \cite{kubo,zubarev}:
\begin{equation}
\Gamma^{erg}_{\psi\psi^{\dagger}}({\bf i},{\bf j})=\langle
\psi({\bf i}) \rangle \langle \psi^{\dagger}({\bf j}) \rangle
\end{equation}

In this manuscript, we focus on the Heisenberg chain of spin-1/2
with periodic boundary conditions in presence of an external
magnetic field directed along $z$ and proportional to $h$:
\begin{equation}
H=J\sum_{{\bf i}}{\vec S}(i){\vec
S}\left(i^\alpha\right)-h\sum_{{\bf i}}S^{z}(i)
\end{equation}
where $i^\alpha=({\bf i}+1,t)$, $J$ is the exchange coupling and
${\vec S}(i)$ is the spin-1/2 operator at the site ${\bf i}$ of
the chain of $N$ sites.

\section{Results}

\begin{figure}[t]
\begin{center}
\epsfxsize=8cm \epsfbox{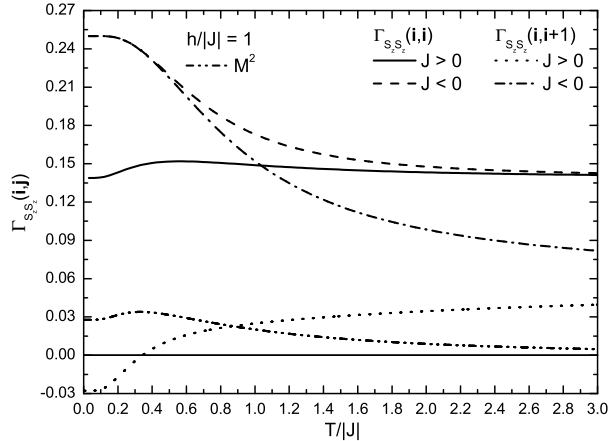}
\end{center}
\caption{For the 3-site chain, the on-site and nearest-neighbor
zero frequency functions, for antiferromagnetic ($J>0$; solid line
and dotted line, respectively) and ferromagnetic ($J<0$; dashed
and dot-dashed line, respectively) exchange coupling, together
with squared magnetization $M^2$ (double dot-dashed line), at
magnetic field $h/|J|=1$, are plotted versus temperature}
\label{Fig1}
\end{figure}

We diagonalize exactly the Heisenberg model on three sites in
magnetic field, obtaining eigenvalues and eigenvectors. Then,
Eq.~(\ref{Gdef}) yields the following functional dependence on
temperature and magnetic field of the zero-frequency function
$\Gamma_{S_{z}S_{z}}({\bf i},{\bf j})$ (see Fig.~\ref{Fig1}):
\begin{eqnarray}
\Gamma_{S_{z}S_{z}}({\bf i},{\bf i})={5\over 36}+\frac{1-2{\rm
e}^{\beta h}+{\rm e}^{2\beta h} }{9\left[1+{\rm e}^{2\beta
h}+2{\rm e}^{{3\over 2}\beta J}{\rm e}^{\beta
h}\right]}\\
\Gamma_{S_{z}S_{z}}({\bf i},{\bf i}+1)=-{1\over 36}+\frac{5-4{\rm
e}^{\beta h}+5{\rm e}^{2\beta h}}{18\left[1+{\rm e}^{2\beta
h}+2{\rm e}^{{3\over 2}\beta J}{\rm e}^{\beta h}\right]}
\end{eqnarray}
We can see that the zero-frequency functions behave quite
differently then their ergodic values $M^2=\langle S_z \rangle^2$
and with a quite strong dependence on both magnetic field and
temperature. According to this, a careful determination of them is
absolutely necessary and we cannot rely at all on the ergodic
values.

Let us now analyze the situation as function of the number $N$ of
sites. To perform calculations for chains with a number of sites
greater than 4 it is necessary to use numerical tools like the
exact diagonalization and the Lanczos method. The thermodynamic
limit will be extrapolated by means of a $1/N$ analysis.

For zero temperature the ergodic value of the on-site
zero-frequency functions $\Gamma_{S^{+}S^{-}}$ is always zero, not
depending on the magnetic field; for the on-site
$\Gamma_{S^{z}S^{z}}$ it is zero excluding three cases: 1) $J<0$
and finite magnetic field ($\Gamma^{erg}_{S^{z}S^{z}}={1\over
4}$); 2) $J>0$ and high magnetic field $h\gg J$
($\Gamma^{erg}_{S^{z}S^{z}}={1\over 4}$); 3) $J>0$ and magnetic
field with values corresponding to energy level crossing (for
finite N; e.g., $h=3J/2$ for 3 sites).

Properly calculated zero-frequency functions $\Gamma_{S^{z}S^{z}}$
and $\Gamma_{S^{+}S^{-}}$ behave differently. For $J<0$, $T=0$ and
$h=0$ both constants are non-zero; this is due to the degeneracy
of the ground state coming from the fact that both $S^{z}$ and
$S^{+}$ are constants of motion and higher value of the total spin
is selected. We also have the exact relation $\Gamma_{S^{z}S^{z}}
= {1\over 2} \Gamma_{S^{+}S^{-}}$, which can be derived from the
algebra in absence of magnetization. The inclusion of a non-zero
magnetic field removes the degeneracy, by fixing the value of
$S^{z}$ and giving dynamics to $S^{+}$, and switches the functions
to their ergodic values.

For $J>0$, $T=0$ and $h=0$ both constants are zero or non-zero,
depending whether the number of sites in the chain is even or odd,
respectively. The relation $\Gamma_{S^{z}S^{z}}={1\over
2}\Gamma_{S^{+}S^{-}}$ still holds as the magnetization is zero in
this case too. The discrepancy with respect to the ergodic values
for {\emph odd} chains is due to the impossibility to pair up all
spins on selecting the lower value of the total spin: {\emph even}
chains have a unique $S^{z}=0$ singlet-like ground state, the
{\emph odd} ones have a degenerate $S^{z}=1/2$ ground state. The
inclusion of the magnetic field sets $\Gamma_{S^{+}S^{-}}$ again
to zero (except for the values generating a level crossing and,
consequently, a degeneracy), while $\Gamma_{S^{z}S^{z}}$ keeps its
previous value or jumps to a higher one passing through level
crossing values. As the magnetic field becomes high enough the
system becomes fully polarized, the ground state is non-degenerate
as for $J<0$ and $h>0$, and both constants set at their ergodic
values.

\begin{figure}[t]
\begin{center}
\epsfxsize=8cm \epsfbox{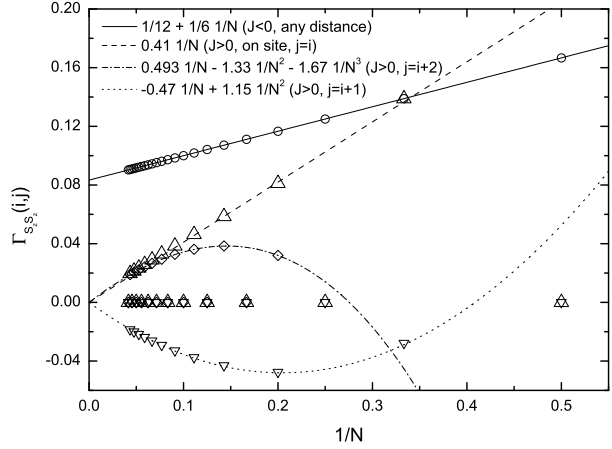}
\end{center}
\caption{Zero-frequency functions for a $N$-site chain with
ferromagnetic ($J<0$; $\bigcirc$) and antiferromagnetic ($J>0$)
coupling. In the latter case, zero-frequency functions are
calculated on-site ($\bigtriangleup$), for first
($\bigtriangledown$) and second neighbors ($\diamondsuit$). The
lines are fits of the numerical data,
%(of odd N only, for $J>0$),
according to the formulae displayed in the legend} \label{Fig2}
\end{figure}

These results were obtained solving numerically chains up to 25
sites, using exact diagonalization and Lanczos method. Now the
main question: do these results hold also in the thermodynamic
limit? In Fig.~\ref{Fig2} we show the zero-frequency function
$\Gamma_{S^{z}S^{z}}$ as a function of the inverse of the number
of sites $1/N$ for $J>0$ and $J<0$, $T=0$ and $h=0$. For $J<0$,
$\Gamma_{S^{z}S^{z}}({\bf i},{\bf j})$ is independent of the
distance $|{\bf i}-{\bf j}|$ and is exactly given by $1/12 + 1/(6
N)$ \cite{note}. For $J>0$, $\Gamma_{S^{z}S^{z}}({\bf i},{\bf j})$
is a function of the distance $|{\bf i}-{\bf j}|$ and scales with
$N$ as shown in the figure (the lines represent very accurate
fits). According to this, we have a non-ergodic behavior for $J<0$
which remains in the thermodynamic limit too. We wish to point out
that this result is exact and must be taken into consideration by
whoever needs to compute physical
quantities like spin correlation functions and susceptibilities in
the 1D Heisenberg model by means of analytical methods.
$\Gamma_{S^{z}S^{z}}$ should be set to $1/12$; any other value,
the null ergodic one too, will induce wrong results and any
approximation scheme not taking this into account, directly or
self-consistently, will surely fail to reproduce the physical
properties of the system. Non-ergodic behaviors for finite values
of temperature and magnetic field cannot be excluded at all and
are instead probable; we are performing numerical calculations in
this direction and the results will be published elsewhere.
Finally, it is worth noting that non-ergodic behaviors with a
functional structure (i.e., zero-frequency functions with
dependencies on temperature and magnetic fields as rich as those
found for finite size chains) require the maximum attention in
building up reliable conserving and/or self-consistent
approximation scheme as any mistake in reproducing them will mine
the possibility to comprehend the low-energy properties of the
system.


\begin{references}
\bibitem[14]{note} In this specific case ($T=0$, $h=0$), the formula can be also exactly obtained by combinatorics.
\end{references}
\end{document}